\documentclass[final]{aipproc}

\layoutstyle{8x11double}

\newcommand{\nat}{Nature}
\newcommand{\sci}{Science}
\newcommand{\apjl}{Astrophys. J.}
\newcommand{\apj}{Astrophys. J.}
\newcommand{\aj}{Astron. J.}
\newcommand{\mnras}{MNRAS}
\newcommand{\aap}{Astron. Astrophys.}
\newcommand{\aaps}{Astron. Astrophys. Suppl. Ser.}
\newcommand{\lesssim}{\mathrel{\hbox{\rlap{\hbox{\lower4pt\hbox{$\sim$}}}\hbox{$<$}}}}
\newcommand{\gtrsim}{\mathrel{\hbox{\rlap{\hbox{\lower4pt\hbox{$\sim$}}}\hbox{$>$}}}}
\newcommand\arcsec{\mbox{$^{\prime\prime}$}}
\newcommand\arcmin{\mbox{$^{\prime}$}}

\begin{document}

\title{Twenty Years of Searching for (and Finding) Globular Cluster
  Pulsars}


\author{Scott M. Ransom}{address={National Radio Astronomy
    Observatory, 520 Edgemont Rd., Charlottesville, VA, 22903 USA}}

\begin{abstract}
  Globular clusters produce orders of magnitude more millisecond
  pulsars per unit mass than the Galactic disk.  Since the first
  cluster pulsar was uncovered twenty years ago, at least 138 have
  been identified -- most of which are binary millisecond pulsars.
  Because of their origins involving stellar encounters, many of these
  systems are exotic objects that would never be observed in the
  Galactic disk.  Examples include pulsar---main sequence binaries,
  extremely rapid rotators (including the current record holder), and
  millisecond pulsars in highly eccentric orbits.  These systems are
  allowing new probes of the interstellar medium, the equation of
  state of material at supra-nuclear density, the mass distribution of
  neutron stars, and the dynamics of globular clusters.
\end{abstract}

\maketitle

\section{Introduction}

The first globular cluster (GC) pulsar was identified 20 years ago in
the cluster M28 after intense efforts by an international team
\cite{lbm+87}.  Since then at least 138 GC pulsars\footnote{For an
  up-to-date catalog of known GC pulsars, see Paulo Freire's website
  at \url{http://www.naic.edu/~pfreire/GCpsr.html}}, the vast majority
of which are millisecond pulsars (MSPs), have been found.  Finding
these GC pulsars has required high-performance computing,
sophisticated algorithms, state-of-the-art instrumentation, and deep
observations with some of the largest radio telescopes in the world,
primarily Parkes, Arecibo, and the Green Bank Telescope (GBT).  The
payoff has been an extraordinarily wide variety of science.

Low-Mass X-ray Binaries (LMXBs) have been known to be
orders-of-magnitude more numerous per unit mass in GCs as compared to
the Galactic disk since the mid-1970s \cite{kat75,cla75}.  This
overabundance is due to the production of compact binary systems
containing primordially-produced neutron stars via stellar
interactions within the high-density cluster cores.  Since LMXBs are
the progenitors of MSPs, this dynamics-driven production mechanism
also applies to them, and it has made GCs (particularly the massive,
dense, and nearby ones) lucrative targets for deep pulsar searches.

Camilo \& Rasio \cite{cr05} produced an excellent review of the first
100 GC pulsars in 2005.  This current review provides a significant
update to Camilo \& Rasio as it concentrates on the advances made
(primarily with the GBT) within the past several years, including
almost 40 additional pulsars and over 50 new timing solutions.

\section{Basic Properties of GC Pulsars}

\begin{figure}
  \includegraphics[height=0.73\textheight,angle=270]{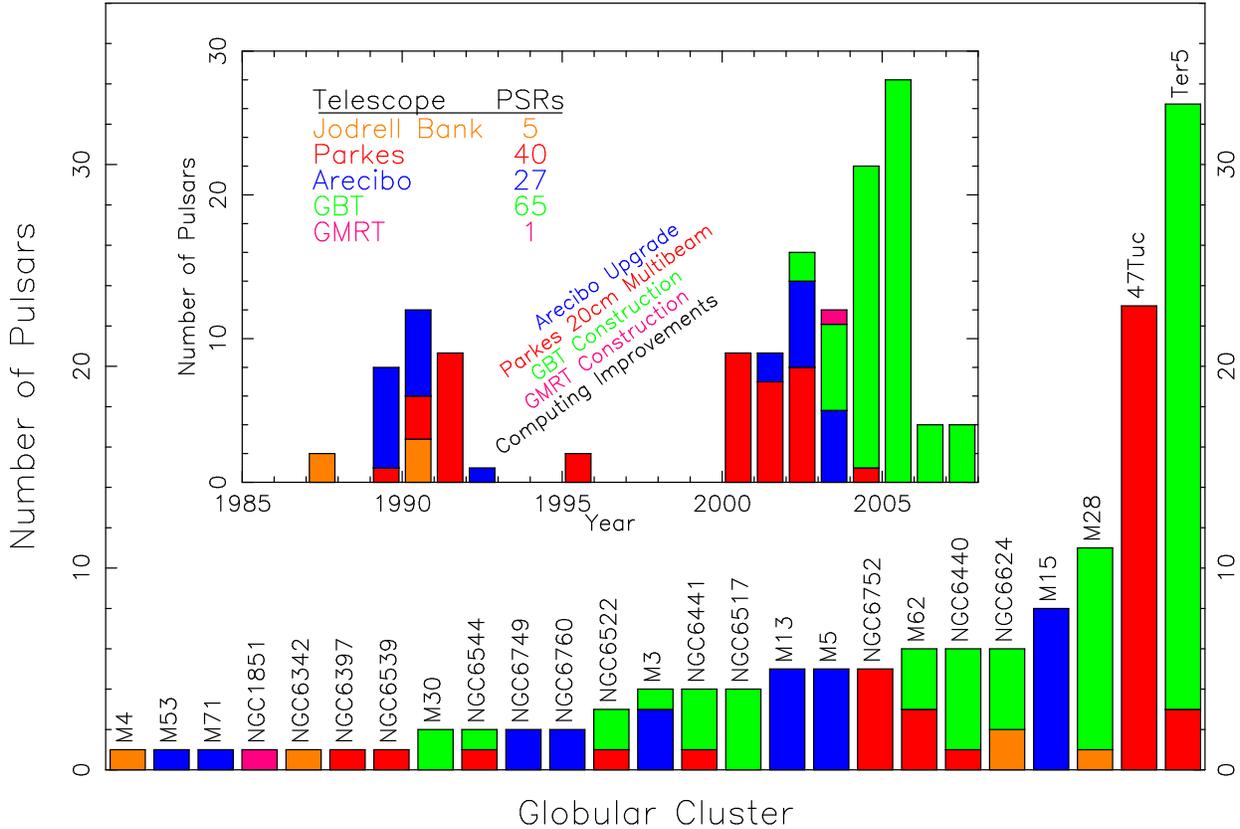}
  \caption{\label{fig:timeline}The number of pulsars per globular
    cluster and \emph{(inset)} a timeline of globular cluster pulsar
    discoveries.}
\end{figure}

There are currently 138 known pulsars in 25 different GCs
\footnote{The Galaxy has roughly 150 known GCs \cite{har96}.}.  Over
100 GC pulsars have been found in the past 10 years, with almost 60 of
these coming in the last 4 years from searches using the GBT (see
Figure~\ref{fig:timeline}).  The three clusters Terzan~5, 47~Tucanae,
and M28 account for approximately half of these pulsars, with 33, 23,
and 11 pulsars in each cluster respectively.  Of the known pulsars, 80
are members of binary systems\footnote{One of the ``binaries'',
  B1620$-$26 aka M4A, is in fact a confirmed triple system including a
  white dwarf and a planet \cite{tacl99,srh+03}.}, 50 are isolated,
and 8 are as yet undetermined.

\begin{figure}
  \includegraphics[height=0.35\textheight,angle=270]{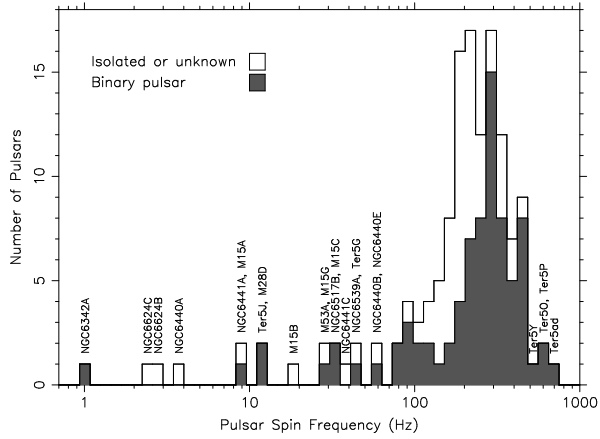}
  \caption{\label{fig:spinfreqs}Spin frequency histogram of the 138
    currently known GC pulsars.  Eighty of the pulsars are confirmed
    members of binaries, 50 are isolated, and 8 are as yet
    undetermined.  It is interesting to note that the binary MSPs seem
    to spin more rapidly on average than the isolated MSPs.  Perhaps
    this is an indication that they are in general younger (i.e. more
    recently recycled) than the isolated MSPs. Such an explanation
    makes sense if all isolated MSPs originally come from binaries and
    therefore must destroy their companions over time.}
\end{figure}

Almost 90\% of GC pulsars are true MSPs with spin periods
$P_{spin}<20$\,ms (see Figure~\ref{fig:spinfreqs}). Their spin
properties seem consistent with the standard ``recycling'' scenario
\cite[e.g.][]{acrs82}, with surface magnetic field strengths of
$B\sim10^{8-9}$\,G and characteristic ages of
$\tau_c\sim10^{9-10}$\,yrs.  Most of the rest of the pulsars are
partially recycled.  However, there are several seemingly very
out-of-place ``normal'' radio pulsars with $P_{spin}>0.2$\,s and
$\tau_c\sim10^{7}$\,yrs as well \cite{lmd96}.

\begin{figure}
  \includegraphics[height=0.35\textheight,angle=270]{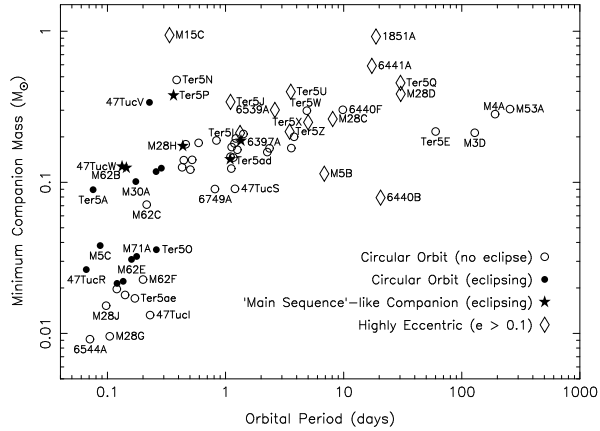}
  \caption{\label{fig:binaries}Orbital period $P_{orb}$ vs. minimum
    companion mass ($M_{c,min}$; assuming a pulsar mass of
    1.4\,M$_\odot$) for the 70 binary GC pulsars with well-determined
    orbits.  All 6 pulsars with ``main sequence''-like companions and
    14 of the known pulsars in highly eccentric ($e > 0.1$) orbits are
    labeled, as well as many more ``normal'' binary systems. Pulsars
    in ``NGC'' clusters are labeled without ``NGC'' to save space.
    The large grouping of pulsars with $P_{orb} < 1$\,day and
    $M_{c,min} < 0.04$\,M$_\odot$ are the so-called ``Black Widow''
    systems. The lack of pulsars in the lower right portion of the
    diagram is not due to selection effects, as those pulsars (if they
    existed) would be relatively easy to identify during searches or
    through timing observations.  M15C is the only confirmed GC double
    neutron star system \cite{jcj+06} while Ter5N is likely the only
    known GC pulsar with a Carbon-Oxygen WD companion \cite{rhs+05}.}
\end{figure}

There are at least four distinct groups of binary GC pulsars (see
Figure~\ref{fig:binaries}).  The first two are similar to the binary
MSPs found in the Galactic disk.  The ``Black Widows'' have very low
mass companions ($M_c\lesssim0.04$\,M$_\odot$) and orbital periods of
several hours, while the ``normal'' low-mass binary MSPs (LMBPs)
likely have Helium white dwarf (WD) companions of mass
$M_c\sim0.1-0.2$\,M$_\odot$ and orbital periods of several to tens of
days. The Black Widow systems are relatively much more common in GCs
($\sim$25\% of the binaries) than in the Galactic disk, though
($\sim$4\% of the binaries).

The other two groups of binary GC pulsars are possibly unique to
clusters and their formation therefore likely depends on the high
stellar densities and interactions found in the cores of GCs.
Approximately 10\% of GC binaries appear to have ``main
sequence''-like companions which show irregular eclipses, erratic
timing, and often have hard X-ray and/or optical counterparts.  The
prototype system is J1740$-$5340 in NGC~6397 \cite{dpm+01}.  Finally,
$\sim$20\% of the known GC binaries have highly eccentric orbits (with
$e > 0.1$).  The standard recycling scenario produces circular orbits
due to tidal interactions during mass transfer, and so the large
eccentricities are probably either induced during multiple stellar
interactions with passing stars \cite[e.g.][]{rh95} or produced
directly during an exchange encounter with another star or binary.
Over the past several years the numbers of pulsars in each of these
two groups have grown dramatically.  This is likely due to the fact
that recent surveys have successfully probed many of the most massive
and dense clusters in the Galaxy where these systems are
preferentially produced (i.e. M28, M62, 47~Tucanae, Terzan~5,
NGC~6440, and NGC~6441).

\section{Searching for Cluster Pulsars}

The signal-to-noise ratio from a radio observation towards a (perhaps
unknown) pulsar is $\propto
S_{\nu}\,A_{e}\,T_{sys}^{-1}\,\sqrt{B_{\nu}\,t_{obs}}$.  $S_{\nu}$ is
the pulsar's flux density at the observing frequency $\nu$.  $A_{e}$
is the effective collecting area of the telescope.  $T_{sys}$ is the
system temperature, which at the $\sim$GHz radio frequencies of
interest is roughly the sum of the receiver temperature (typically
15$-$25\,K), the 2.7\,K cosmic microwave background, and the Galactic
synchrotron background ($T_{Gal}$$\propto$$\nu^{-2.6}$ and typically a
few to tens of K depending on sky position and observing frequency
$\nu$).  $B_{\nu}$ is the radio bandwidth used for the observation and
$t_{obs}$ is the observation duration.  In addition, signal-to-noise
ratios improve when a pulsar has short duration pulsations compared to
its pulse period (i.e. a small pulse duty cycle).  If the pulses are
smeared or broadened in time, perhaps by uncorrected orbital motion or
interstellar medium effects, signal-to-noise ratios during searches
can be reduced to effectively zero\footnote{See Lorimer \& Kramer
  \cite{lk04} for a much more thorough discussion of pulsar search
  sensitivities.}.

\subsection{The Problems with Cluster Distances}

MSPs are intrinsically very faint radio sources.  Because of this, all
of the wide-area Galactic pulsar surveys conducted to date have been
severely sensitivity limited for MSPs (many have been instrumentation
limited as well).  Of the $\sim$60 known Galactic MSPs, $\sim$80\% are
within 2\,kpc of the Sun.  In contrast, the \emph{nearest} GCs (M4 and
NGC~6397) are just over 2\,kpc from the Sun, and most, including the
best targets for pulsar searches, are at distances of 5$-$15\,kpc
\cite{har96}.

In addition to the inverse square law problem, the large distances to
GCs often (especially for the bulge clusters near the Galactic center)
imply large column densities of the interstellar medium (ISM).  The
ionized ISM causes frequency-dependent dispersion of radio waves
($\propto\nu^{-2}$), scatter-broadening of the radio pulses
($\propto\nu^{-4.4}$), and for certain clusters dramatic fluctuations
of observed pulse intensity due to diffractive scintillation
\cite[e.g.][]{clf+00}.  These ISM effects, as well as a substantially
reduced Galactic synchrotron background, have pushed typical observing
frequencies from $\sim$400\,MHz in the early 1990s up to 1.3$-$2\,GHz
in the past decade.  At these frequencies, especially with much wider
observing bandwidths available (hundreds of MHz), significant
sensitivity gains have been realized despite the usually steep radio
spectra of the pulsars themselves (flux densities
$S_{\nu}\propto\nu^{\alpha}$ with -3$\lesssim\alpha\lesssim$-1 and
$<$$\alpha$$>$$\sim$$-$1.8; \cite{mkk+00}).

\subsection{Searching for Binaries}

Since most GC pulsars are in binaries\footnote{The current binary
  fraction of $\sim$60\% is a lower limit since finding isolated
  pulsars is much easier than finding binaries.}, orbital motion
causes Doppler variations of the observed pulsation frequencies during
an observation.  If uncorrected, these variations can make even very
bright MSPs undetectable.  However, correcting for unknown orbital
motion identically would be extremely computationally expensive and so
current searches only account for linear changes in apparent spin
frequencies (i.e. constant $\dot f$).  These ``acceleration'' searches
\cite[e.g.][]{jk91} are valid when the orbital period is much longer
than the observation duration ($P_{orb}$$\gtrsim$10\,$t_{obs}$).

Acceleration searches add an extra dimension to the traditionally
two-dimensional phase space of dispersion measure DM\footnote{DM is
  the integrated electron column density along the line-of-sight to a
  pulsar.} and spin frequency $f$ over which one must search for
pulsars.  Since the number of trials in the acceleration or $\dot f$
dimension is proportional to $t_{obs}^2$, the long observations used
to improve the sensitivity of GC searches greatly increase their
computational costs.  Typically, searches for the first pulsar in a
cluster are made using a large range of likely DMs but only a limited
range of accelerations or $\dot f$.  Once the first pulsar is found
and the rough DM toward the cluster is known, a much smaller range of
DMs is searched, but with a much larger range of possible
accelerations for additional pulsars.

As an example, to properly search a single 7\,hr GBT observation of
Terzan~5 (where the DM is known to $\sim$5\%) with a full range of
acceleration searches, requires approximately one CPU-\emph{year} of
processing on state-of-the-art CPUs. However, it is important to
realize that without acceleration searches (or other advanced binary
search techniques such as Dynamic Power Spectra \cite{cha03}), the
majority of the binary GC MSPs that have been uncovered over the past
decade would simply not have been found.

\subsection{A Renaissance in 2000}

The above paragraphs summarize why GC pulsar searches require long
integrations (sensitivity) at GHz frequencies (minimize ISM effects),
using the largest telescopes (collecting area), the best receivers
(wide bandwidths and low system noise), and large amounts of
high-performance computing (acceleration searches).  The dramatic
increases in the numbers of known GC pulsars beginning in 2000 (see
Figure~\ref{fig:timeline}) resulted from significant improvements in
each of these areas.  First, new low-noise and wide-bandwidth
($B_{\nu}$$\sim$300\,MHz) observing systems centered near 1.4\,GHz
became available at Parkes and Arecibo.  Second, the rise of
affordable cluster-computing allowed acceleration searches to be
conducted at investigator institutions rather than at special
supercomputing sites.  The first major success from these improvements
(and a significant driver for further efforts) was the discovery of 9
new binary MSPs in 47~Tuc \cite{clf+00}.

The third and perhaps most important improvement was the completion of
the GBT in 2001.  With its state-of-the-art receivers, approximately
three times greater $A_e$ than Parkes, and the ability to observe over
80\% of the celestial sphere, it is perfectly suited to make deep
observations of GCs.  By the end of 2003, a fantastic wide-bandwidth
($B_{\nu}$$\sim$600\,MHz) system centered near 2\,GHz became available
which provided 5$-$20 times more sensitivity for MSPs in certain GCs
in the Galactic bulge than the 1.4\,GHz system used at Parkes.  The
discovery of 30 new MSPs in Terzan~5 \cite{rhs+05}, a cluster
previously extensively searched at Parkes, including the fastest known
MSP (J1748$-$2446ad aka Ter5ad; \cite{hrs+06}), were some of the first
results.  Pulsar surveys of many additional GCs using the same system
are ongoing.

\subsection{Future Cluster Pulsar Surveys}

Recent work on the luminosities $L$ of GC pulsars \cite{hrs+07} has
confirmed earlier results \cite[e.g.][]{and92} suggesting that the
luminosity distribution roughly follows a $d\,\log\,N = -d\,\log\,L$
relation.  In addition, this work implies that we currently observe
only the most luminous pulsars in each cluster.  Together, these facts
indicate that our current GC pulsar surveys are completely sensitivity
limited such that even marginal improvements in search sensitivities
will result in new pulsars\footnote{47~Tuc is a possible exception to
  this rule as deep radio imaging \cite{mdca04} and X-ray observations
  \cite{gch+02} indicate that scintillation may have already allowed
  the identification of nearly all of the observable MSPs in the
  cluster.}.  The history of GC pulsar searches has directly
demonstrated this fact many times.

However, there seems to be little likelihood of making very large
improvements in GC pulsar search sensitivities over the next several
years.  Most of the variables in the signal-to-noise equation are
already nearly optimal (e.g. $T_{sys}$, $B_{\nu}$, $t_{obs}$, and
$\nu$).  Dramatic improvements in sensitivities and therefore pulsar
numbers will almost certainly require a new generation of larger
telescopes (i.e. larger $A_e$) such as FAST\footnote{FAST:
  \url{http://www.bao.ac.cn/LT/}} or the SKA\footnote{SKA:
  \url{http://www.skatelescope.org}}.

\section{Which clusters have pulsars?}

Figure~\ref{fig:timeline} shows the number of pulsars in each of the
GCs with known pulsars.  Currently there are 10 clusters with 5 or
more pulsars and 3 clusters with 10 or more pulsars: M28 with 11,
47~Tucanae with 23, and Terzan~5 with 33.  Camilo \& Rasio \cite{cr05}
pointed out that there are very few clear correlations between cluster
parameters and the numbers of known pulsars.  In fact, the only simple
properties that seem to be related to the number of known pulsars are
the total mass of the cluster (which likely influenced how many
neutron stars were originally retained) and the distance $D$ to the
cluster (since all GC pulsar searches are currently sensitivity
limited).  However, even these indicators have exceptions.  For
example, $\omega$ Centauri has been searched extensively but
unsuccessfully with the Parkes telescope, yet it is one of the nearest
and most massive GCs in the Galactic system.

A more sophisticated indicator of which clusters may contain more
LMXBs and therefore MSPs is the predicted stellar interaction rate
$\Gamma_c$ in the cores of the clusters, where LMXBs and MSPs are
likely formed.  Pooley et al.~\cite{pla+03} showed a strong
correlation between the number of X-ray sources in a cluster and its
$\Gamma_c$, which they expressed as $\Gamma_c \propto \rho_0^{1.5}
r_c^2$, where $\rho_0$ is the central density and $r_c$ is the core
radius.  We can attempt to adjust the indicator to account for our
senstivity issues by ranking clusters by $\Gamma_c\,D^{-2}$.  Using
this metric, we find that many of the clusters with numerous pulsars
are near the top of the list, including 47~Tuc, Terzan~5, M62,
NGC~6440, NGC~6441, NGC~6544, M28, and M15.  Also near the top are
several others which likely contain numerous pulsars but whose
positions near or behind the Galactic center region (and therefore
large amounts of ISM) make searches very difficult (e.g. NGC~6388 and
Liller~1 \cite{fg00}).

It is important to realize, though, that because of limited amounts of
telescope time, pulsar searchers have specifically \emph{targeted}
those clusters near the top of the $\Gamma_c\,D^{-2}$ list first.
Therefore, clusters further down that list may simply not have known
pulsars because they haven't been searched to the same sensitivity
levels as the clusters near the top of the list.  Until we have a
large number of GCs, independent of their position on the
$\Gamma_c\,D^{-2}$ list, searched to sensitivities comparable to the
recent GBT 2\,GHz surveys, it will be difficult to determine just how
good of a predictor $\Gamma_c\,D^{-2}$ really is.

The recent 1.4\,GHz survey of all 22 GCs within 50\,kpc and visible
with Arecibo \cite{hrs+07} is a good example of the type of surveys we
need.  The Arecibo survey found 11 new MSPs, the majority of which are
in clusters with fairly average values of $\Gamma_c\,D^{-2}$.  No
pulsars were found (or have ever been found) in clusters with very low
central luminosity densities, $\rho_0 < 10^3$\,L$_\odot$\,pc$^{-3}$.
A similar unbiased survey of $\sim$60 GCs at 1.4\,GHz using the Parkes
telescope has uncovered 12 new pulsars \cite{dpm+01,pdm+01} in 6 GCs.
Unfortunately, the limited sensitivity of that survey does not rule
out even relatively bright MSPs in many of the clusters, thereby
making it difficult to draw conclusions about cluster properties and
their pulsar populations.

\subsection{What pulsars are in those clusters?}

We can compare the pulsar populations in the best studied clusters,
such as Terzan~5 and 47~Tucanae, to attempt to determine if the
properties of the clusters affect their pulsars.  Two of the simplest
things to compare are the spin-period distributions and the binary
populations, both of which do show significant differences.

The spin-period distribution of the Terzan~5 pulsars is significantly
flatter than that of the 47~Tuc pulsars with more faster and more
slower pulsars.  In fact, Ter~5 contains 5 of the 10 fastest known
spinning pulsars \cite{hrs+06}, while 47~Tuc contains only one within
the top ten.  Likewise, 47~Tuc has no MSPs rotating slower than
$\sim$8\,ms, whereas Ter~5 has six.  A Kolmogorov-Smirnov test
suggests a $<$10\% chance that the two period distributions were drawn
from the same parent distribution.  As for the binary populations,
Ter~5 has only two known ``Black-Widow'' systems compared to five in
47~Tuc.  On the other hand, Ter~5 has six highly-eccentric binaries
compared to none in 47~Tuc.  Are these differences related to the
current interaction rates in the cluster cores (Ter~5's is 2$-$3 times
that of 47~Tuc's) or perhaps the ``epochs'' when the MSP production
rates were the highest?

The pulsars in M28, NGC~6440, and NGC~6441 are more similar to those
in Ter~5 than to those in 47~Tuc, both in terms of their spin periods
(i.e. flatter distributions) and their binary parameters (with a
broader mix of different systems).  It is interesting to note, though,
that the 10 known pulsars in the very similar clusters NGC~6440 and
NGC~6441 rotate on average significantly slower than those in either
Ter~5 or 47~Tuc. Only one pulsar rotates faster than 5\,ms, and half
rotate slower than 13\,ms \cite{frb+07}.

\section{Timing of Cluster Pulsars}

While it is obviously essential to \emph{find} the pulsars in GCs to
do any science with them, without detailed follow-up observations, and
in particular pulsar timing solutions, the amount of science one can
do is severely limited.  For most GC pulsars the extraordinary
precision provided by MSP timing provides $\sim$0.1\arcsec\ and often
significantly better astrometric positions (crucial for
multi-wavelength follow-up \cite[e.g.][]{gch+02}), extremely precise
Keplerian orbital parameters for binaries, and measurements of the
pulsar's apparent spin period derivative.  For some pulsars,
particularly those in eccentric orbits, certain post-Keplerian orbital
parameters can be measured \cite[e.g.][]{lk04} which allow the
determination of (or at least constraints on) the masses of the pulsar
and/or the companion star.  Establishing timing solutions for as many
GC pulsars as possible allows us to use them both individually and in
ensembles to probe a wide variety of both basic physics and
astrophysical phenomena.  Over the past two years, the number of GC
pulsars with timing solutions (currently 107) has almost doubled,
resulting in many interesting new results.

\subsection{Ensembles of pulsars} 

\begin{figure}
  \includegraphics[height=0.53\textheight,clip=True]{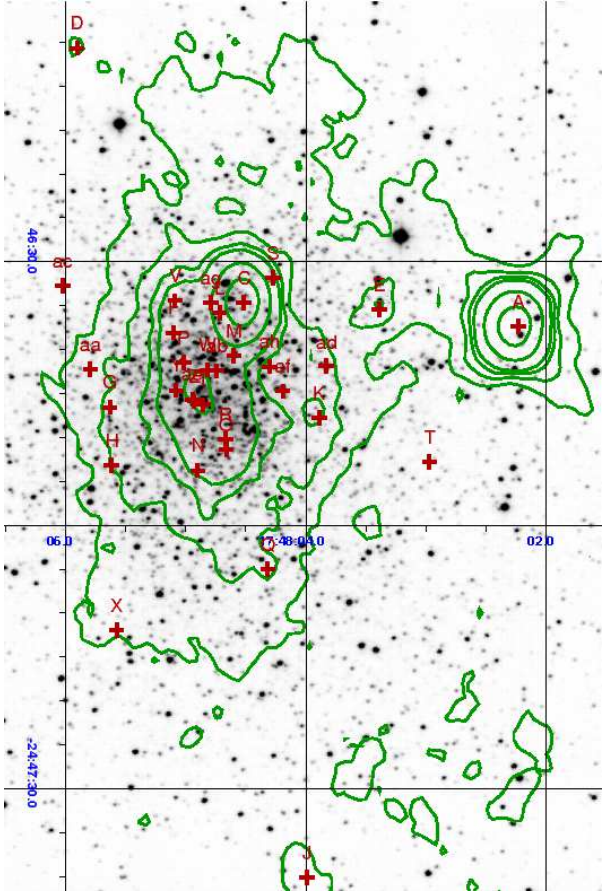}
  \caption{\label{fig:Ter5}Positions for the 32 Terzan~5 pulsars with
    timing solutions overlaid on 1.4\,GHz radio contours from the VLA
    \cite{fg00}, and an NTT I-band image (courtesy S.~Ortolani;
    \cite{obb96}).  The average positional error in RA is
    $\lesssim$0.02\arcsec\ while for DEC it is $\lesssim$0.2\arcsec\
    (the much larger DEC errors are due to the fact that Terzan~5 is
    very near the ecliptic). The good correspondence between the
    timing positions and the VLA radio contours implies that
    potentially all of the radio flux is indeed produced by unresolved
    radio pulsars (many tens of which are as yet unknown).}
\end{figure}

The largest ensemble of pulsars in a single GC with timing solutions
are the 32 in Terzan~5 (only Ter5U, a weak eccentric binary, remains
without a solution; see Figure~\ref{fig:Ter5}).  However, there are
five other GCs with at least five pulsars with timing solutions as
well (47~Tuc \cite{fck+03}, M28 \cite{beg06}, M15 \cite{and92},
NGC~6440 \cite{frb+07}, and NGC~6752 \cite{cpl+06}).  These ensembles
of pulsars in individual GCs can produce unique science.

\begin{figure}
  \includegraphics[height=0.35\textheight,angle=270]{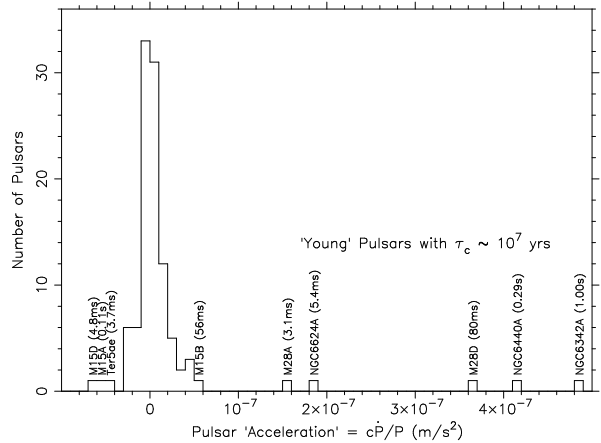}
  \caption{\label{fig:accels}Histogram of observed globular cluster
    pulsar ``accelerations'' ($c\dot P_{obs}/P_{obs}$). For GC
    pulsars, the observed acceleration is the sum of the pulsar's
    intrinsic acceleration (i.e. $c\dot P_{int}/P_{int}$), an apparent
    acceleration due to the pulsar's proper motion, and accelerations
    from both the Galaxy's and the globular cluster's gravitational
    potentials \cite{phi92,phi93}.  The proper motion and Galactic
    potential terms are typically small compared to the others.  Since
    GC pulsars are typically fully recycled with small intrinsic
    accelerations ($\sim10^{-9}$\,m\,s$^{-2}$), the GC gravitational
    accelerations ($\sim|10^{-8}|$\,m\,s$^{-2}$) usually dominate.
    This is apparent in the figure by a nearly symmetric clustering of
    pulsars around zero acceleration.  The pulsars on the observer's
    side of a cluster receive positive accelerations while those on
    the far side receive negative accelerations and appear to spin
    more rapidly with time.  The five pulsars with observed
    accelerations $>10^{-7}$\,m\,s$^{-2}$ are anomalously young
    (characteristic ages
    $1\times10^7\lesssim\tau_c\lesssim3\times10^7$\,years) and their
    intrinsic accelerations are much larger than the maximum possible
    gravitational acceleration from their clusters.  Apparently
    clusters continue to produce such systems \cite[e.g.][]{ihr+07}.
    The three pulsars with the most negative accelerations provide
    unique probes of the central dynamics (and lower limits on the
    mass-to-light ratio) of their parent clusters
    \cite[e.g.][]{and92}.}
\end{figure}

\paragraph{Probes of Ionized ISM and Intra-Cluster Medium}
The precise DMs (errors $<$0.1\,pc\,cm$^{-3}$) and timing positions
for 32 pulsars in Ter~5 have recently allowed a unique probe of the
ionized Galactic ISM between us and the cluster on parsec scales
\cite{ran07}.  A calculation of the DM structure function indicates
that the fluctuations in the ISM on 0.2$-$2\,pc scales roughly follow
those predicted for Kolmogorov turbulence.  Earlier work on 47~Tuc
using the pulsar positions, accelerations (see
Figure~\ref{fig:accels}), and DMs provided the first definitive
measurement of ionized gas within a GC \cite{fkl+01}.

\begin{figure}
  \includegraphics[height=0.35\textheight,angle=270]{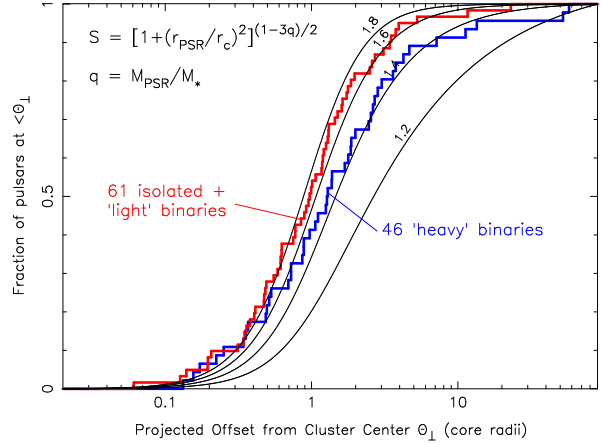}
  \caption{\label{fig:radial}Radial distribution comparison of
    ``isolated'' pulsars (including binaries with very low-mass
    companions, $M_c \lesssim 0.05$\,M$_\odot$) and heavier binary
    pulsar systems.  The numbers on the thin black lines are various
    $q = M_p/M_\star$ values where $M_p$ is the mass of the pulsar
    \emph{system} and $M_\star$ is the mass of the dominant stellar
    component in the GCs ($M_\star \sim 0.9$\,M$_\odot$).
    Surprisingly, the ``isolated'' systems appear to be more centrally
    condensed (and therefore possibly more massive) than the binary
    systems.}
\end{figure}

\paragraph{Statistical Neutron Star Mass Measurement}
Using the projected offsets of pulsars from their cluster centers and
a model for how relaxed components of different masses should be
distributed within a cluster, it is possible to statistically measure
the masses of the pulsar systems $M_p$ \cite{hgl+03}.
Figure~\ref{fig:radial} shows the 107 GC pulsars with timing positions
split into two different groups: 1) isolated pulsars or binary pulsars
with very low-mass companions ($M_{c,min} \lesssim 0.05$\,M$_\odot$)
and 2) binary pulsars with more massive companions.  Surprisingly, the
``isolated'' systems seem to be more centrally condensed than the
(supposedly more massive) binary systems.  Fits of the observed
distributions give $q = M_p / M_\star \sim 1.5$ for the binaries and
$q \sim 1.7$ for the ``isolated'' systems.  Assuming that the dominant
stellar components in the cluster cores have mass $M_\star =
0.9$\,M$_\odot$ implies binary system masses of $M_{p,bin} \sim
1.35$\,M$_\odot$ and \emph{larger} masses $M_{p,iso} \sim
1.53$\,M$_\odot$ for the ``isolated'' pulsars.  The reason for this
difference in mass segregation is currently unknown.

\paragraph{Cluster Proper Motions}
The very precise positions available from MSP timing allow individual
pulsar proper motions given regular observations over 5$-$10 years.
The measurement of several pulsar proper motions from a single cluster
allows a measurement of the proper motion of the GC itself.  Currently
this has been accomplished for three clusters (47~Tuc \cite{fck+03},
M15 \cite{jcj+06}, and NGC~6752 \cite{cpl+06}), and several more will
likely be measured within the next couple of years.  Such measurements
are very important for determining the Galactic orbits of GCs and
predicting the effects of tidal stripping and/or destruction.
Measuring cluster proper motions is very difficult in the optical
(using {\em HST}, for instance), especially for the Galactic bulge
clusters which are distant and plagued by extinction.

\paragraph{Cluster Dynamics}
The projected positions of the pulsars with respect to the cluster
centers as well as measurements of their period derivatives (which are
usually dominated by acceleration within the gravitational potential
of the GC; see Figure~\ref{fig:accels}) provide a sensitive probe into
the dynamics of the cluster and even constrain the mass-to-light ratio
near the cluster center \cite{phi92}.  These measurements can provide
evidence for the presence (or absence) of black holes in the cores of
the clusters \cite{dpf+02}.  Cluster dynamics also influences pulsars
by ejecting some of them to the outskirts of the clusters or even
entirely \cite{ihr+07}.  A recently uncovered example of such a system
is M28F, a bright (for a GC MSP) isolated pulsar located almost
3\arcmin\ from the center of M28.  That offset is larger than for any
other GC pulsar except for NGC~6752A \cite{cpl+06}, where exotic
ejection mechanisms have been invoked to explain its position
\cite[e.g.][]{cmp03}.

\subsection{Individual Exotic Pulsars}

While ensemble studies of GC pulsars are very interesting, many of the
pulsars are truly \emph{exotic} or unique objects, worth studying
individually.

\paragraph{Young and Slow Pulsars}
A handful of slow, ``normal'' ($\tau_c\sim10^7$\,yrs) pulsars have
been known in GCs for some time: B1718$-$19 aka NGC~6342A
\cite{lbhb93}, B1820$-$30B aka NGC~6624B \cite{bbl+94}, and B1745$-$20
aka NGC~6440A \cite{lmd96}.  Recently, at least one more slow pulsar
has been uncovered, NGC~6624C with $P_{spin}$ = 0.405\,s \cite{cha03},
as well as M28D, an 80\,ms binary pulsar that is definitely ``young''
(see Figure~\ref{fig:accels}).  The slow pulsars, which have likely
not been through any recycling, must have formed relatively recently
even though all of the massive stars would have gone supernova
$10^{10}$\,yrs ago.  One possibility is that the pulsars formed via
electron-capture supernovae, perhaps via accretion-induced collapse of
a massive WD or merger-induced collapse of coalescing double WDs
\cite[see N.~Ivanova's paper in these proceedings;][]{ihr+07}.

\paragraph{The fastest MSPs}
Ter5ad is the fastest MSP known, with a spin period of $P_{spin}$ =
1.396\,ms \cite{hrs+06}.  Its discovery finally broke the 23-yr-old
``speed'' record established by the very first MSP discovered
\cite{bkh+82} and renewed hope for finding a sub-MSP (a pulsar having
a spin period under 1\,ms).  A sub-MSP would provide by far the most
direct and interesting constraints on the properties of matter at
nuclear densities and would be of major significance to physics in
general \cite{lp07}.  Besides Ter5ad, several other very rapid
rotators have been uncovered in Ter5 (Ter5O at 1.676\,ms and Ter5P at
1.728\,ms; \cite{rhs+05}), and it seems likely that the first sub-MSP
(if they exist) might be found in a GC.

\paragraph{``Main-Sequence''$-$MSP Systems}
Several pulsars have been recently discovered (including 47TucW
\cite{egc+02,bgv05}, M62B \cite{pdm+03}, Ter5P \cite{rhs+05}, Ter5ad
\cite{hrs+06}, and M28H \cite{beg06}) which appear to have bloated
``main-sequence''-like companion stars much like the prototype system
J1740$-$5340 in NGC~6397 \cite{dpm+01,fpds01}.  These pulsars are
eclipsed for large fractions of their orbits and show irregular
eclipses on some occasions. Timing positions usually associate them
with hard X-ray point sources where the high-energy emission is likely
generated via colliding MSP and companion winds.  Several of the
pulsars have been identified in the optical where they exhibit
variability at the orbital period.  In addition, at least some of them
exhibit highly erratic orbital variability (resulting in several large
amplitude orbital period derivatives) likely due to tidal interactions
with the bloated companion stars.  These systems could be the result
of an exchange encounter between a main sequence star and a ``normal''
binary MSP system.  Alternatively, perhaps the companions are the
stars that have recently recycled the pulsars and we are observing
newly born MSPs \cite{dpm+01,fpds01}.  Multi-wavelength studies of
these systems are difficult \cite[e.g.][]{bgv05}, but allow a wide
variety of additional constraints to be placed on MSP emission
mechanisms, their winds, and the evolutionary histories of their
systems (including tidal circularization theory).

\paragraph{Highly Eccentric Binaries}
At least 15 GC pulsar systems are members of eccentric binaries with
$e > 0.1$, and most of those contain MSPs.  In contrast, only a single
eccentric binary MSP is known in the Galactic disk\footnote{See
  D.~Champion's contribution to these proceedings.}.  Ten of these
systems have been discovered since 2004: six in Terzan~5
\cite{rhs+05}, two in M28 \cite{beg06}, and one each in NGC~1851
\cite{fgri04,frg07} and NGC~6440 \cite{frb+07}.  Given the angular
reference that ellipses provide, pulsar timing can easily measure the
precession of the angle of periastron or $\dot\omega$.  For compact
companions, $\dot\omega$ will be dominated by general relativistic
effects, and its measurement provides the total mass of the binary
system \cite{lk04}.  Timing observations of four of these systems
(Ter5I \& J \cite{rhs+05}, NGC~6440B \cite{frb+07}, and M5B [Freire et
al. in prep.]) indicate ``massive'' neutron stars
($>$1.7\,M$_{\odot}$) which constrain the equation-of-state of matter
at nuclear densities \cite{lp07}.  Such constraints are impossible to
achieve in nuclear physics laboratories here on Earth.  In addition,
similar measurements for M28C, a 4.15\,ms pulsar in an 8-day orbit
with $e=0.85$, indicate that the pulsar is \emph{less} massive than
1.37\,M$_\odot$.  This is a fairly low mass for a neutron star which
must have accreted a substantial amount of material during recycling
and will likely constrain recycling models.  For more information on
these systems, see P.~Freire's contribution to these proceedings.

\paragraph{Other Exotica}
There is already one confirmed GC triple system (PSR~B1620$-$26 in M4)
which contains an MSP, a white dwarf, and a planetary-mass component
\cite{tacl99,srh+03}.  Intriguingly, ongoing observations show very
strange and seemingly systematic timing residuals from the
``isolated'' MSP NGC~6440C \cite{beg06}.  One explanation for these
residuals is the presence of one or more terrestrial-mass planets.
Given the strange variety of systems that have already been found in
GCs, it is quite possible that one of the many currently
uncharacterized systems could be another unique object.

\section{Prospects for the Future}

Given the wide variety of science (most of which was unanticipated)
that has already resulted from GC pulsars, and the fact that we are
currently only seeing a small fraction (perhaps 10-20\%) of the total
GC pulsar population, the future of GC pulsar astronomy seems very
bright.  Improvements in search sensitivities with current instruments
will likely uncover tens of additional systems and next-generation
telescopes like FAST or the SKA promise to find hundreds.  Among these
pulsars may be even more spectacular ``exotica'' such as MSP$-$MSP or
MSP$-$black hole binaries, and from these new pulsars will come many
surprising results.

\begin{theacknowledgments}
  Thanks go to my collaborators on the many recent search efforts made
  with Arecibo and the GBT: Steve B\'egin, Ryan Lynch, Jennifer Katz,
  Lucy Frey, Mike McCarty, Ben Sulman, Fernando Camilo, Vicky Kaspi,
  and especially Jason Hessels, Ingrid Stairs, and Paulo Freire.
\end{theacknowledgments}

\bibliographystyle{aipproc}

\end{document}